\documentstyle[12pt,epsf]{article}
\renewcommand{\bar}[1]{\overline{#1}}

\begin{document}

\begin{flushright}
USM-TH-93
%hep-ph/00xxxx
\end{flushright}

\bigskip\bigskip
\centerline{\large {\bf Nuclear Effects on the Extraction of
Neutron Structure Functions }}

\vspace{22pt}

\centerline{\bf  Ivan Schmidt\footnote{e-mail:
ischmidt@fis.utfsm.cl} $^{a}$, Jian-Jun Yang\footnote{e-mail:
jjyang@fis.utfsm.cl}$^{a,b}$}

\vspace{8pt}

{\centerline {$^{a}$Departamento de F\'\i sica, Universidad
T\'ecnica Federico Santa Mar\'\i a,}}

{\centerline {Casilla 110-V, %\\ \null\quad
Valpara\'\i so, Chile\footnote{Mailing address}}}

{\centerline {$^{b}$Department of Physics, Nanjing Normal
University,}}

{\centerline {Nanjing 210097, China}}

%\vfill

\vspace{10pt}
\begin{center} {\large \bf Abstract}
\end{center}

Nuclear effects in light nuclei due to the presence of spin-one
isosinglet 6-quark clusters are investigated. The quark
distributions of 6-quark clusters are obtained by using a
perturbative QCD (pQCD) based framework, which allows us to get a
good description of the ratio of the deuteron structure function
to the free nucleon structure function. Nuclear effects on the
extraction of the neutron structure functions $F_2^n$ and $g_1^n$
are estimated. We find that the effect on the extracted
spin-dependent neutron structure function is very different from
that on  the spin-independent neutron structure function. The
effect  enhances  the  Bjorken sum by about 10\%, whereas its
correction to the Gottfried sum is rather small. The formalism for
calculating nuclear effects is further used to evaluate the
spin-dependent structure function of the $^3$He nucleus  and a
good self-consistent check is obtained.

\vfill

\centerline{PACS numbers: 21.60.Gx, 25.30.Mr, 24.85.+p, 14.20.Dh}

%\vfill
%\centerline{Submitted to Phys.~Rev.~C/D. for publication}
\vfill
\newpage

\section{Introduction}

It is well known that nuclear effects are commonly extracted from
the analysis of structure functions ratios of different nuclei
relative to the deuteron in the belief that they are negligible in
this last system. On the other hand, some years ago the New Muon
Collaboration(NMC)~\cite{NMCPRD94,NMCPRL91} reported a value of
the Gottfried sum, using the ratio $F^n_2/F^p_2=2F^D_2/F_2^p-1$,
where nuclear effects in the deuteron target were neglected, and
its structure function was regarded as the sum of the structure
functions of the proton and neutron. Then the neutron structure
function $F_2^n$ can be extracted, and the ratio of the deuteron
structure function $F^D_2$ to the free nucleon structure function
$F_2^N=(F^p_2+F^n_2)/2$ should read $R_{F}^{D/N}=1$. However,
Gomez et al.~\cite{Gomez} found that the deuteron has a
significant EMC effect~\cite{EMC}, i.e. $R_{F}^{D/N} \not= 1$,
especially in the region near $x \sim 0.6$. The analysis of Epele
et al.~\cite{Epele} also shows a significant nuclear effect due to
the composite nature of the deuteron. In fact, the importance of
nuclear effects on the extraction of the neutron structure
functions was recognized even much earlier~\cite{Schmidt}. Recent
investigations of nuclear effects on a deuteron mainly emphasize
the following three aspects: (1) the EMC effect on the
spin-independent deuteron structure function~\cite{Gomez,Burov};
(2) the effects of Fermi motion of nucleons in the
deuteron~\cite{Braun,Sidorov,Tokarev}; (3) depolarization of the
nucleons and spin-dependent effects~\cite{Benesh}. In recent work
by Burov and Molochkov~\cite{Burov}, the EMC effect on a deuteron
is analyzed from a relativistic point of view. Fermi motion for
the spin-independent $F_2^D$ and $F_3^D$ structure functions in
light-cone variables was analyzed in Refs.~\cite{Braun}
and~\cite{Sidorov}, respectively. The same relativistic approach
and a deuteron model have been used~\cite{Tokarev} in order to 
describe the effect of Fermi motion corrections to the 
spin-dependent structure function $g_1^D$, and to estimate 
nuclear effects in the $ \mu + D \to \mu + X$ process. But there 
have been no detailed studies of the possible nuclear effects on 
the spin-dependent quark distributions themselves and the effects
on the extraction of the neutron structure functions ($F_2^n, 
g^n_1$). Nuclear effects including Fermi motion and shadowing  
are very important in order to extract the neutron deep-inelastic
structure functions  from the experimental data for deuteron and 
heavy nuclei. They should be included into any QCD analysis of 
the nucleon structure functions. The main purpose of our present 
work is to investigate nuclear effects on the extraction of the 
neutron structure functions.

Generally, it is  hard to give a unified  description of all
nuclear effects including Fermi motion corrections at large $x$,
and shadowing and  anti-shadowing  effects in the small $x$
region. The work by Lassila and Sukhatme~\cite{Lassila} shows that
the quark cluster  model (QCM) can be used to model the EMC effect
over all $x$. The QCM has been successfully applied to inclusive
electron deuteron reactions at large momentum transfer, and to
nuclear Drell-Yan processes. Spin-dependent effects in the
quark-cluster model of the deuteron and $^3$He were investigated
by Benesh and Vary~\cite{Benesh}. However, in Ref.~\cite{Benesh}
there were no  detailed 6-quark clusters quark distributions, and
hence only the first moment of a particular distribution was
involved in the estimation. In order to get more precise
information of nuclear effects on the extraction of the neutron
structure functions, detailed quark distributions of 6-quark
clusters are necessary. This motivates us to construct the quark
distributions of 6-quark clusters. Several years ago, Brodsky,
Burkardt and Schmidt provided a reasonable description of the
spin-dependent quark distributions of the nucleon in a pQCD based
model~\cite{Bro95}. This model has also been successfully used in
order to explain the large single-spin asymmetries found in
semi-inclusive pion production in $pp$ collisions, while other
models have not been able to fit the data~\cite{Bog99}. Recently
it has been also applied to obtain a good description of the spin
and flavor structure of octet baryons~\cite{MSY4}, specially the
$\Lambda$ particle~\cite{MSY2,MSY3,MSSY5}. In this paper, we
extend this analysis to 6-quark clusters, trying to describe its
spin-dependent quark distributions in this framework.

The spin-dependent structure functions are interesting not only
because they introduce a new physical variable with which to
explore the detailed structure of the nucleon, but also because
they provide a precise test of QCD via the Bjorken sum rule, which
is a strict QCD prediction. In this paper, we will investigate
nuclear effects on the extraction of the spin-dependent  neutron
structure function by considering the nuclear effects in the
deuteron.  We will see that nuclear effects cause a big
enhancement of the Bjorken sum  by about 10\%, giving a final 
result which is much closer to the theoretical value of the 
Bjorken sum rule.

The paper is organized as follows. In section 2 we will describe
the quark distributions of a 6-quark cluster in the pQCD based
model. In section 3 we estimate  the nuclear effects due to the
presence of 6-quark clusters in the deuteron, and on the
corresponding extraction of the spin-independent neutron structure
function $F_2^n$. The ratio of the deuteron structure function to
the free nucleon structure function and the effect on the
Gottfried sum are calculated. We find that our model can give a
very good unified description of the nuclear effects in the
deuteron structure function and that the effect on the Gottfried
sum is very small. In section 4 we present an analysis of the
possible effect on the extraction of the spin-dependent neutron
structure function and the Bjorken sum. A significant effect on
the spin-dependent neutron structure function, and therefore on
the Bjorken sum, is obtained. The formalism for calculation the
nuclear effect is checked by applying it to the $^3$He nucleus.
Finally, a discussion and summary are included in section 5.

\section{Quark Distributions in Spin-One Isosinglet 6-Quark Cluster}

In order to describe the spin-dependent quark distributions of the
nucleon,  Brodsky, Burkardt and Schmidt developed a pQCD based
model~\cite{Bro95}. Here we extend this analysis to the
description of the quark distributions in a 6-quark cluster. In
the region $z \to 1$, where $z$ is the light-cone momentum
fraction carried by a given quark or antiquark in the 6-quark
cluster ($0 \leq z \leq 1$), pQCD can give rigorous predictions
for the behavior of distribution functions. In particular, it
predicts ``helicity retention", which means that the helicity of a
valence quark will match that of the parent quark cluster.
Explicitly the quark distributions of a spin-one isosinglet
6-quark cluster  satisfy the counting rule~\cite{countingr},
\begin{equation}
Q_6(z) \sim (1-z)^p, \label{pl}
\end{equation}
where
\begin{equation}
p=2 n-1 +2 \Delta S_z.
\end{equation}
Here $n$ is the minimal number of the spectator quarks, and
$\Delta S_z=|S_z^q-S_z^6|=1/2$ or $3/2$ for parallel or
anti-parallel quark and the 6-quark cluster  helicities,
respectively. More specifically,  spin distributions for the
non-strange quarks in the  6-quark cluster  can be parameterized
as,

\begin{equation}
Q^{\uparrow}_6(z)=\frac{1}{z^{\alpha}} [
A_{6}(1-z)^{10}+B_{6}(1-z)^{11} ], \label{Qup}
\end{equation}

\begin{equation}
Q^{\downarrow}_6(z)=\frac{1}{z^{\alpha}} [
C_{6}(1-z)^{12}+D_{6}(1-z)^{13} ].  \label{Qdw}
\end{equation}
Here the distributions include both the quark and anti-quark
contributions, i.e. $ Q=q + \bar{q} $, and $q$ is regarded  as the
sum of the non-strange quarks, i.e. $q=u+d$, since there are no
detailed experimental data guiding us to determine separate quark
distributions in the 6-quark cluster. The effective QCD Pomeron
intercept $\alpha = 1.12 $ is introduced to reflect the Regge
behavior at low $z$ and is chosen to have the same value as that
for the nucleon~\cite{Bro95}. There are 4 parameters for the above
quark distributions. In consideration of the limiting case in
which the spin-independent and spin-dependent structure functions
of the 6-quark cluster become $F_2^6=(F_2^p+F_2^n)$ and
$g_1^6=(g_1^p+g_1^n)$ respectively, we make the assumption that
the value of the helicity carried by the quarks in the 6-quark
cluster is twice as large as that in the nucleon. With this
assumption, the first moment of the spin-dependent structure
function of the 6-quark cluster is  compatible with that obtained
by $s$-wave MIT bag wave functions~\cite{Benesh}. Therefore we fix
the parameters by using the following conditions: one condition
arises from the requirement that the sum rules converge at $z \to
0$; the second condition from the values of the integral of the
polarized quark distribution $\Delta Q_6 =0.68$ which is $2
(\Delta u +\Delta d)$ of the nucleon~\cite{Stiegler}; the third
condition reflects the fact that the momentum fraction $z_Q$
carried by the quark and anti-quark of a 6-quark cluster should be
half of that for a single nucleon, i.e.,
$z_Q=\frac{1}{2}(x_u+x_d)=0.2605$~\cite{Martin}. This leaves us
with one unknown, which is chosen to be $C_6$. The three
constraints give the solution set

\begin{equation}
\begin{array}{clllc}
A_{6}=~~1.0203 C_{6}+3.3455,
\\
B_{6}=-1.0954 C_{6}-2.3519,
\\
D_{6}=-1.0751 C_{6}+0.9936.
\end{array}
\label{abdqs}
\end{equation}
The probabilistic interpretation of parton distributions
$Q^{\uparrow}_6$ and $Q^{\downarrow}_6$ implies the bounds
\begin{equation}
0< C_{6} < 13.225.
\end{equation}

Similarly,  the strange quark and antiquark distributions are 
parameterized as

\begin{equation}
S^{\uparrow}_6(z)=\frac{1}{z^{\alpha}} [
A_{s}(1-z)^{12}+B_{s}(1-z)^{13} ], \label{Sup}
\end{equation}

\begin{equation}
S^{\downarrow}_6(z)=\frac{1}{z^{\alpha}} [
C_{s}(1-z)^{14}+D_{s}(1-z)^{15} ],  \label{Sdw}
\end{equation}
with
\begin{equation}
\begin{array}{clllc}
A_{s}=~~1.0145 C_{s} - 2.4230,
\\
B_{s}=-1.0785 C_{s} + 2.5966,
\\
D_{s}=-1.0640 C_{s} + 0.1736,
\end{array}
\label{abdqs2}
\end{equation}
which are constrained by (1) the requirement that the sum rules
converge at $z \to 0$; (2) the values of $\Delta S_6 =-0.24$,
twice as large as the value $\Delta S=-0.12$ for the
nucleon~\cite{Stiegler}; (3) the sum of momentum fractions carried
by the strange quarks and antiquarks, $z_s=0.0175$, half of the
value $x_s=0.035$ for the nucleon~\cite{Bro95}. The probabilistic
interpretation of parton distributions $S^{\uparrow}_6$ and
$S^{\downarrow}_6$ requires

\begin{equation}
 2.389 < C_{s} < 2.713.
\end{equation}

In the following calculation, first we consider only the
contributions of non-strange quarks  and  determine the value
$C_{6}=9.5$ by fitting the data of the ratio of the deuteron
structure function to the free nucleon structure function. Then we
investigate the effect due to the presence of the strange quark.

\section{Nuclear Effects on Extraction of Spin-independent Neutron  Structure
Function}

Nuclear effects on the extracted neutron structure function will
be indicated by the  ratio $R^{D/N}_{F}$ of the deuteron structure
function to the free nucleon structure function, and the
modification of the Gottfried sum.

\subsection{Ratio $R^{D/N}_{F}$}

The EMC effect indicates that there is  overlap of nucleons and
the existence of 6-quark clusters in a nucleus. Now we will use
the above 6-quark cluster quark distributions in order to
understand the nuclear effects in the deuteron structure function.

The probability for creating a 6-quark cluster in the deuteron has
been calculated as $p_6=0.054$~\cite{Benesh}. We use a simple
model where a fraction $p_6$ of the nucleons in deuterium have
become part of a 6-quark cluster, with a fraction $(1-p_6)$
remaining single nucleons. In order to compare with the
experimental data, we introduce the ratio

\begin{equation}
R_{F}^{D/N}=\frac{F^D_2}{\tilde{F}^N_2},
\end{equation}
where

\begin{equation}
\tilde{F}^N_2=\frac{1}{2} (F^p_2+\tilde{F}^n_2),
\end{equation}
with

\begin{equation}
\tilde{F}^n_2(x)=\frac{\left[2 F_2^D(x)-p_6
\tilde{F}_2^6(\frac{x}{2}) \right]}{1-p_6} -F_2^p(x), \label{Fnb}
\end{equation}
assuming that the clusters formed in the $s$-wave and $d$-wave
components have the same structure.  We denote the neutron
structure function as $\tilde{F}^n_2$, which is different from
$F^n_2$, extracted via
\begin{equation}
F^n_2(x)=2F_2^D(x) -F_2^p(x), \label{Fn}
\end{equation}
without considering nuclear effects. In (\ref{Fnb}),
the structure function of the 6-quark cluster can be expressed as

\begin{equation}
\tilde{F}^6_2(z)=\frac{z}{9}[5 Q_6(z) + 2 S_6(z)],
\end{equation}
where
\begin{equation}
Q_6(z)=Q_6^{\uparrow}(z)+Q_6^{\downarrow}(z),
\end{equation}
and
\begin{equation}
S_6(z)=S^{\uparrow}_6(z)+S^{\downarrow}_6(z).
\end{equation}
In Eqs.~(\ref{Fnb}) and (\ref{Fn}), the deuteron structure
function $F_2^D$ and the proton structure function $F_2^p$ are
take from the parametrizations of NMC~\cite{NMCPLB95}. Our
numerical calculations show that the effect of the strange quark
is very small. In order to understand the data of $R_{F}^{D/N}$ at
$Q^2=4 GeV^2$, the one free parameter for the non-strange quark
distribution $Q_6$ of the 6-quark cluster should be chosen as
$C_6=9.5$. In Fig.~1 we present the calculated results of the
ratio $R^{D/N}_{F}$, together with the data which were extracted 
by Gomez et al.~\cite{Gomez} using a model  of Frankfurt and
Strikman~\cite{Frankfurt}. We obtain a good unified description of
the nuclear effects including shadowing, anti-shadowing at small
$x$ and Fermi motion at large $x$, with the presence of 6-quark
clusters. It has been commonly accepted that the overlap of the
nucleons in nuclei can result in the nuclear shadowing due to the
parton recombination at small $x$~\cite{Close,Kumano,Yang}. The
quarks in a 6-quark cluster have a larger confinement size than
those in the nucleon, which leads to the re-distribution of the
quarks towards to  the lower momentum end, i.e. lower $x$ region
according to the uncertainty principle. This causes a suppression
of the quark distributions in the medium $x$ region and an
enhancement in the low $x$ region. The competition of the
enhancement effect with the shadowing effect leads to the
anti-shadowing effect at $x\sim 0.1$. A rise of the ratio at large
$x$ is expected from Fermi motion of the nucleons, and this is
related to the nuclear wave function for overlapping nucleons
which is here modeled as a 6-quark cluster. The nuclear effects
over all $x$ have been unified modeled as the presence of the
6-quark clusters in the deuteron.  The strange quark in the
6-quark clusters only gives very small modification, and there is
no significant change in its effect when the free parameter $C_s$
varies in the admissible range [2.389, 2.713]. For simplicity, we
only show the strange contribution with $C_s=2.4$ in the following
results.

\begin{figure}[htb]
\begin{center}
\leavevmode {\epsfysize=5.5cm \epsffile{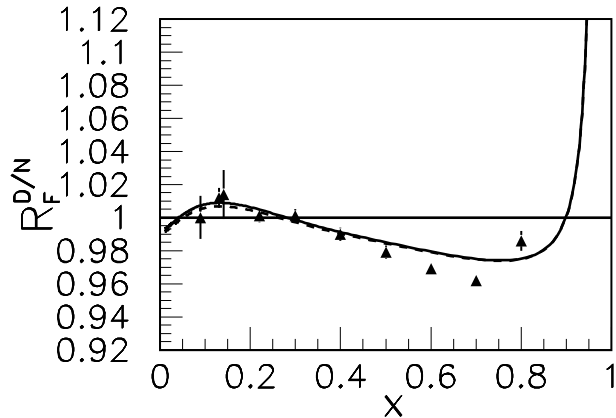}}
\end{center}
\caption[*]{\baselineskip 13pt The ratio $R_{F}^{D/N}(x)$ and the
comparison with the data from Ref.~\cite{Gomez}. Dashed and solid
curves correspond to including  only contributions of non-strange
quarks, and including also  the  contribution of the strange quark
in the 6-quark cluster, respectively. }\label{sy3f1}
\end{figure}

\subsection{The Effect on the Gottfried Sum}

Recently, the New Muon Collaboration (NMC)
experiment~\cite{NMCPRD94} has provided values for the ratio of
the structure function $F^n_2/F^p_2$, assuming that nuclear
effects are not significant in the deuteron, i.e.,

\begin{equation}
F_2^p-F^n_2=2 F_2^D \frac{1-F_2^n/F_2^p}{1+F^n_2/F^p_2},
\end{equation}
where

\begin{equation}
\frac{F_2^n}{F_2^p}=2 \frac{F_2^D}{F_2^p}-1.
\end{equation}
The Gottfried sum is defined as~\cite{GS}

\begin{equation}
S_{GS}=\int S_{GF}(x) dx,
\end{equation}
with the Gottfried integrand function given by
\begin{equation}
S_{GF}(x)=\frac{(F^p_2(x)-F^n_2(x))}{x}.
\end{equation}
 When we allow for the presence of 6-quark clusters in the
deuteron, the neutron structure function should be extracted as
$\tilde{F}_2^n$ from Eq.~(\ref{Fnb}). The corresponding Gottfried
integrand should be modified from $S_{GF}$ to

\begin{equation}
\tilde{S}_{GF}(x)=\frac{(F^p_2(x)-\tilde{F}^n_2(x))}{x}.
\end{equation}
In order to visualize the nuclear effects, the ratio

\begin{equation}
R_{GF}(x)=\frac{\tilde{S}_{GF}(x)}{S_{GF}(x)}
\end{equation}
is shown in Fig.~2. Taking the minimal value of $x$ as
$x_{min}=0.004$, which corresponds to the kinematic limit of the
experiment in Ref.~\cite{NMCPRL91}, we obtain the  value of the
Gottfried sum,

\begin{equation}
 S_{GS}=\int\limits_{0.004}^1 S_{GF}(x) dx=0.226,
\end{equation}
and

\begin{equation}
\tilde{S}_{GS}=\int\limits_{0.004}^1 \tilde{S}_{GF}(x) dx=0.215,
\end{equation}
without taking into account the contribution of the strange quarks
in the 6-quark clusters. We find an approximate  5\% suppression
in the Gottfried sum due to nuclear effects. When the contribution
of the strange quark is also included with $C_s=2.4$, we find that
the Gottfried sum becomes $\tilde{S}_{GS}=0.221$, which is very
close to the value of $S_{GS}$.

In addition, the ratio of the structure function of the neutron to
that of the proton, as shown in Fig.~3, also indicates a weak
modification of the extracted spin-independent neutron structure 
function due to nuclear effects. Therefore nuclear effects on the
Gottfried sum  are very small, although $R_{GF}(x)$ deviates 
obviously from unity as shown in Fig.~2.  The $x \to 1$ behavior 
of the neutron to proton structure function ratio is an important
result which can differentiate between different nucleon 
structure  models~\cite{Thomas}. For example, in exact SU(6) its 
value is 2/3, in some diquark models it is 1/4 ~\cite{Ma}, while 
pQCD predicts 3/7~\cite{Farrar,Bro95}. Our analysis favors the 
pQCD prediction.

\begin{figure}[htb]
\begin{center}
\leavevmode {\epsfysize=5.5cm \epsffile{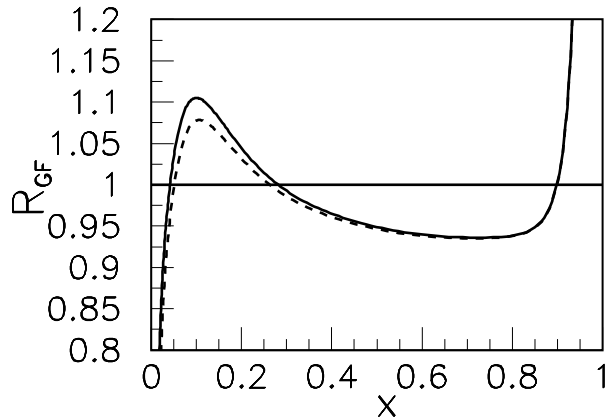}}
\end{center}
\caption[*]{\baselineskip 13pt The ratio $R_{GF}(x)$. Dashed and
solid curves correspond to including only contributions of
non-strange quarks, and including also the contribution of the
strange quark in 6-quark clusters, respectively. }\label{sy3f2}
\end{figure}

\begin{figure}[htb]
\begin{center}
\leavevmode {\epsfysize=5.5cm \epsffile{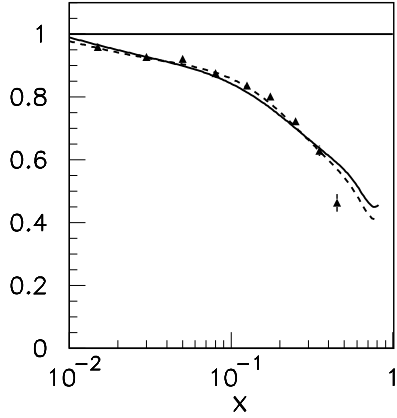}}
\end{center}
\caption[*]{\baselineskip 13pt The ratios   $F_2^n/F_2^p$ and
$\tilde{F}_2^n/F_2^p$ are shown in dashed and solid curves,
respectively. The neutron structure function is extracted from the
deuteron structure function $F_2^D$ and the proton structure
$F_2^p$ which are take from the parametrizations of
NMC~\cite{NMCPLB95}. Experimental data are taken from
Ref.~\cite{NMCPRL91}.}\label{sy3f3}
\end{figure}

\section{Nuclear Effects on Extraction of Spin-dependent
Neutron Structure Function}

Similar to the spin-independent case, we will consider the same
nuclear effects on the extraction of the neutron spin-dependent
structure function. We will indicate these effects by presenting
the result of the ratio $R_{g}^{D/N}$ of deuteron spin-dependent
structure function to the free nucleon spin-dependent structure
function, and the correction  to the Bjorken sum.

\subsection{Ratio $R_{g}^{D/N}$}

In the absence  of 6-quark clusters, only the depolarization of
the proton and neutron in the $d$ wave component of the deuteron
wave function will be taken into account, and hence the
spin-dependent structure function can be written as

\begin{equation}
g_1^D(x)=(p_s-\frac{1}{2} p_d) \frac{[g_1^p(x)
+g_1^n(x)]}{2},\label{g1D}
\end{equation}
where  $p_s$ and $p_d$ denote the probabilities for finding the
deuteron in an $s$ or $d$ wave, respectively. If we allow for the
existence of  6-quark clusters in deuterium, $g_1^D(x)$
becomes~\cite{Benesh}

\begin{equation}
g_1^D(x)=[(p_s-p_{6s})-\frac{1}{2} (p_d-p_{6d})] \frac{[g_1^p(x)
+\tilde{g}_1^n(x)]}{2}+\frac{1}{2} p_6 g^6_1(\frac{x}{2}),
\label{g1Db}
\end{equation}
where $p_{6s}=0.047$ and $p_{6d}=0.007$, the probabilities for
creating a 6-quark cluster in the $s$ and $d$ states, were
calculated by Benesh and Vary~\cite{Benesh}, $p_6=p_{6s}+p_{6d}$,
and $g_1^6$ is the spin-dependent structure function of a spin
one, isosinglet 6-quark cluster,

\begin{equation}
g_1^6(z)=\frac{1}{18}[5 \Delta Q_6(z) + 2 \Delta S_6(z)].
\end{equation}
where

\begin{equation}
\Delta Q_6(z)=Q_6^{\uparrow}(z)-Q_6^{\downarrow}(z),
\end{equation}
and

\begin{equation}
\Delta S_6(z)=S^{\uparrow}_6(z)-S^{\downarrow}_6(z).
\end{equation}
 From Eqs. (\ref{g1D}) and  (\ref{g1Db}), both $g_1^n$, 
$\tilde{g}_1^n$ can be extracted, without 6-quark clusters and 
with 6-quark clusters, respectively. The first moment of the 
spin-dependent structure function of the 6-quark cluster reads

\begin{equation}
\bar{g}^6_1=\int_0^1 g_1^6 (z) d z = 0.189
\end{equation}
including only contributions of non-strange quarks. When the
strange quark is also included, $\bar{g}^6_1$ becomes 0.162, which
is compatible with that obtained by s-wave MIT bag wave
functions\cite{Benesh}.

Similar to the spin-independent case, the nuclear effect is
described by the ratio

\begin{equation}
R_{g}^{D/N}(x)=g_1^D(x)/\tilde{g}_1^N(x)
\end{equation}
and
\begin{equation}
\tilde{g}_1^N(x)=\frac{1}{2}[g_1^p(x)+\tilde{g}_1^n(x)]
\end{equation}
In our numerical calculation, $g_1^D$ is taken from a fit to the
E155 data~\cite{E155PLB99}, and $g_1^p$ from a fit  to the E143
data~\cite{E143PRL95}. The ratio $R_{g}^{D/N}$, as shown in
Fig.~4, is very different from the ratio $R_{F}^{D/N}$ (see
Fig.~1), and it is also very different from that obtained in
Ref.~\cite{Tokarev}, where only Fermi motion corrections were
considered, and whose result is independent of $x$ over a wide 
range of $x=10^{-3} - 0.7$, with a value of 0.9.

\begin{figure}[htb]
\begin{center}
\leavevmode {\epsfysize=5.5cm \epsffile{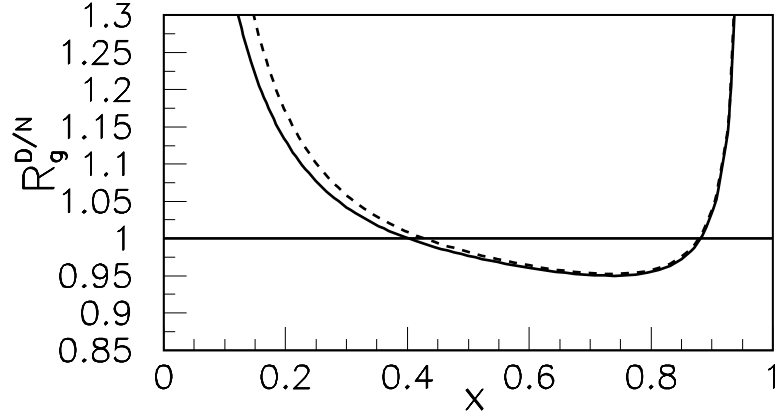}}
\end{center}
\caption[*]{\baselineskip 13pt The ratio $R_{g}^{D/N}(x)$. Dashed
and solid curves correspond to including only contributions of
non-strange quarks, and including also the  contribution of the
strange quark in the 6-quark cluster, respectively. }\label{sy3f4}
\end{figure}

\subsection{The Effect on the Bjorken Sum}

Now, we turn to investigate the nuclear effect on the Bjorken
sum~\cite{BS}. This sum is defined as

\begin{equation}
S_{BS}=\int S_{BF}(x) dx,
\end{equation}
with the Bjorken integrand function

\begin{equation}
S_{BF}(x)=g_1^p(x)-g_1^n(x).
\end{equation}
There is also a  Bjorken sum $\tilde{S}_{BS}$, and a Bjorken
integrand function $\tilde{S}_{BF}(x)$, corresponding to the
extracted $\tilde{g}_1^n$ which includes nuclear effects.

The Bjorken sum rule with perturbative QCD correction to first
order of $\alpha_s$ reads

\begin{equation}
S_{BS}^{th}=\frac{1}{6} g_A [1-\frac{\alpha_s(Q^2)}{\pi}].
\end{equation}
With the well measured neutron beta decay coupling constant
$g_A=1.2601 \pm 0.0025$~\cite{BarnettPRD96} and  very recently
determined QCD coupling $\alpha_s$\cite{Bethke} at $Q^2=4
GeV^2$, one finds

\begin{equation}
S_{BS}^{th} \simeq 0.189. \label{BSth}
\end{equation}

In order to show the nuclear effect on the Bjorken integrand, we
introduce  the ratio

\begin{equation}
R_{BF}(x)=\frac{\tilde{S}_{BF}(x)}{S_{BF}(x)},
\end{equation}
whose value is  larger than unity over the whole $x$ region (see
Fig.~5), therefore indicating an  enhancement of the value of the
Bjorken sum. Actually, the value of the sum  changes from
$S_{BS}=0.166$ to $\tilde{S}_{BS}=0.185$, with an increase of
$\sim 11\%$ when only the contribution of non-strange quarks in
the 6-quark cluster is considered. When the strange quark in the
6-quark cluster is further included with $C_s=2.4$, the Bjorken
sum result decreases to 0.182. This value is very close to the
value in Eq.~(\ref{BSth}), the strict QCD prediction  of the
Bjorken sum rule. Therefore, the nuclear effect due to the
presence of 6-quark cluster favors the case in which the Bjorken
sum rule holds.

\begin{figure}[htb]
\begin{center}
\leavevmode {\epsfysize=5.5cm \epsffile{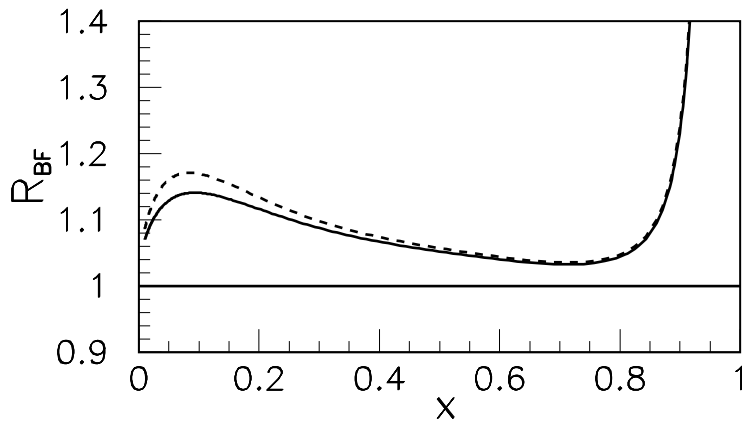}}
\end{center}
\caption[*]{\baselineskip 13pt The ratio $R_{BF}(x)$. Dashed and
solid curves correspond to including only contributions of
non-strange quarks, and including also the contribution of the
strange quark in the 6-quark cluster, respectively. }\label{sy3f5}
\end{figure}

\subsection{The Helium Spin-dependent Structure Function $g_1^{\rm{He}}$}

The $^3$He nucleus is a suitable system in which to check the
above formalism for calculating the nuclear effect.  Recently,
some measurements on $g_1^n$ have been made by using deep
inelastic scattering of polarized electrons on polarized
$^3$He~\cite{E154PRL97,HERMES97, E142PRD96}. In the simplest
picture of $^3$He, all nucleons are in an $S$ wave, which
consequently has a completely anti-symmetric spin-isospin wave
function. The protons in $^3$He are restricted by the Pauli
principle to be in a spin-singlet state. It is usually believed
that a polarized $^3$He nucleus automatically provides a highly
polarized neutron which is rather loosely bound. We can use $^3$He
to have a self-consistent check for the extracted neutron
spin-dependent structure function. With the probabilities of
6-quark clusters in $^3$He calculated by Benesh and Vary with the
Bonn deuteron wave functions~\cite{Benesh}, we assume that the
change of the nuclear environment from Deuterium to Helium does
not alter  the quark spin structure of the 6-quark clusters and
employ the extracted neutron spin-dependent structure function to
predict the spin-dependent structure function of $^3$He. In
Fig.~6,  the spin-dependent structure functions of the neutron and
Helium  are shown in (a) and (b) respectively,  without the
6-quark clusters (dashed curve) and with the nuclear effect due to
the presence of the 6-quark clusters (solid curve). We find that
the calculated spin-dependent structure function of $^3$He almost
does not change, and therefore still provides a good fit to the
experimental data (see Fig.~6(b)), although the neutron
spin-dependent structure function has a significant modification
(see Fig.~6(a)). The contribution to the spin-dependent structure
function of $^3$He from the correction of the neutron
spin-dependent structure function is almost canceled by that from
6-quark clusters. It means that we can also extract the same
neutron spin-dependent structure function from the experimental
data of $g_1^{\rm{He}}$ with the existence of 6-quark clusters in
Helium as that from $g_1^{D}$. This provides a good
self-consistent check of our present framework.

\begin{figure}[htb]
\begin{center}
\leavevmode {\epsfysize=5.5cm \epsffile{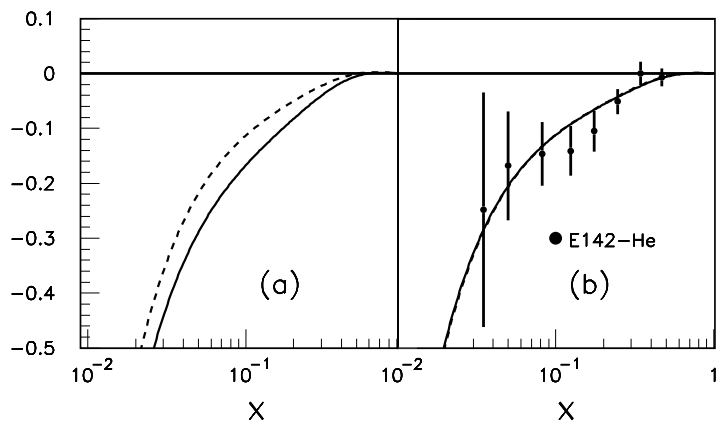}}
\end{center}
\caption[*]{\baselineskip 13pt (a) $g_1^n(x)$ and
$\tilde{g}_1^n(x)$; (b) $g_1^{\rm{He}}(x)$ (dashed curve) and
$\tilde{g}_1^{\rm{He}}(x)$ (solid curve).  The dashed and solid
curves correspond to without considering the 6-quark clusters and
including the nuclear effect due to 6-quark clusters,
respectively. Note that the dashed and solid  curves in (b) almost
overlap. The experimental data  are taken from the
E142~\cite{E142PRD96}. }\label{sy3f6}
\end{figure}

\section{Discussion and Summary}

For simplicity, only 6-quark clusters have been included in our
approach. Inclusion of 9-quark clusters causes the maximum at
$x\simeq 0.1$ to shift a little bit in Fig.~1. The calculations by
Sato~\cite{Sato} show that the probability for forming 9-quark
clusters increases with the nuclear mass number A. In our present
discussion of light nuclei, they can be neglected.

In our numerical calculation,  $Q^2=4 GeV^2$ was chosen in order
to compare  the results of the Gottfried sum with the experimental
data of NMC~\cite{NMCPRD94}. This is also consistent with the
kinematic range of the experiment in which the ratio
$R_{F}^{D/N}(x)$ was extracted~\cite{Gomez}. In principle the 
present simple analytic representations of the quark 
distributions in 6-quark clusters only reflect the intrinsic 
bound-state structure of the 6-quark clusters and they are valid 
at low $Q^2$ where QCD evolution can be neglected. At high $Q^2$,
the radiation from the struck quark line increases the effective 
power-law fall-off $(1-x)^p$ of the structure functions relative 
to the underlying quark distributions: $\Delta p = (4 C_F/\beta_1
) \rm{log}[\rm{log} (Q^2/\Lambda^2)/
\rm{log}(Q_0^2/\Lambda^2)]$~\cite{Bro95}, where $C_F=4/3$ and
$\beta_1=11-\frac{2}{3} n_f$. With this estimate, the effect due
to  the scale change from $Q_0^2=1 GeV^2$ to $Q^2=4 GeV^2$ on our
analysis is small. The quark distributions of the 6-quark cluster
can be used as the input distribution for perturbative QCD
evolution from $Q_0^2$ to a  higher resolution scale.

To sum up, we investigated the nuclear effects in light nuclei due
the presence of 6-quark clusters. The quark distributions of  the
6-quark clusters were modeled in the  pQCD based framework. With
the presence of the 6-quark clusters in the deuteron, the nuclear
effects including shadowing, anti-shadowing and Fermi motion,
which are described by the ratio of the deuteron structure
function to the free nucleon structure function, can be well
explained. The nuclear effects on the Gottfried sum are small. 
Then we extended the formalism for calculating the nuclear 
effects to the extraction of the spin-dependent neutron structure
function. We find that the nuclear effect on the extraction of 
the spin-dependent neutron structure function is more significant
than that on the spin-independent neutron structure function. The
effect results in an increase in  the Bjorken sum  by about 10\%,
which favors the strict QCD prediction of the Bjorken sum rule. A
good self consistent check of the formalism for calculating the
nuclear effects was provided by evaluating the spin-dependent
structure function of the $^3$He nucleus.

 {\bf
Acknowledgments: } This work is partially supported by Fondecyt
(Chile) postdoctoral fellowship 3990048, by Fondecyt (Chile) grant
1990806 and by a C\'atedra Presidencial (Chile), and by National
Natural Science Foundation of China under Grant Numbers 19875024.

\end{document}